\documentclass[twoside]{article}
\usepackage{amsmath}
\usepackage{graphicx}
\usepackage{dcolumn}
\usepackage{color}
\begin{document}
\thispagestyle{plain}
\begin{center}
{\Large \bf \strut
Towards Minkowski space solutions of Dyson--Schwinger Equations through un-Wick rotation
\strut}\\
\vspace{10mm}
{\large \bf 
  Tobias Frederico$^a$,
  Dyana C. Duarte$^a$,    
  Wayne de Paula$^a$,
  Emanuel Ydrefors$^a$,
  Shaoyang Jia$^b$,
  and Pieter Maris$^b$}
\end{center}

\noindent{
{\small $^a$\it  Instituto Tecnol\'ogico da Aeron\'autica, DCTA, 12.228-900 S\~ao Jos\'e dos Campos,  Brazil} 
\small $^b$\it Department of Physics and Astronomy, Iowa State University, Ames, IA 50011, USA} \\
\markboth{
T. Frederico et al. }
{
Dyson--Schwinger Equations through un-Wick rotation} 

\begin{abstract}
The fermion self-energy is calculated from the rainbow-ladder
truncation of the Dyson--Schwinger equation (DSE) in quantum
electrodynamics (QED) for spacelike momenta and in the complex
momentum plane close to the timelike region, both using Pauli--Villars
regularization.  Specifically, the DSE is solved in the complex
momentum plane by rotating either the energy component of the
four-momentum or the magnitude of Euclidean four-momentum to reach the
timelike region in Minkowski space.  The coupling constant is
appropriately chosen to ensure the singularities of the fermion
propagator are located in the timelike region while producing
significant differences from the perturbative solutions.  For
simplicity, we choose Feynman gauge, but the method is applicable in
other covariant gauges as well.  We demonstrate that the approximate
spectral representation based on the fermion self-energy near the
timelike region is consistent with the solution of the DSE directly in
the Euclidean space.  \\[\baselineskip]

\end{abstract}

\section{Motivation}
The measurable quantities associated with the structure of a hadron
state in the full possible kinematical range, which would be obtained
by solving, e.g.  quantum chromodynamics (QCD), require the knowledge
of matrix elements of physical operators with timelike momenta.  This
poses a challenge to methods based on a purely Euclidean formulation
of QCD, using either discretization methods such as lattice gauge
theories, or continuum methods like the Dyson--Schwinger (DSE) and
Bethe-Salpeter equations (BSE)~\cite{Eichmann:2016yit}.
To extract physical observables defined in Minkowski space, these
methods have to rely on an analytic continuation from Euclidean space
such that e.g. the momenta of physical hadrons are on-shell (in the
timelike region).  This is straightforward to do for mesons as bound
states of a quark and anti-quark~\cite{Maris:1997tm,Maris:1999nt}, and can
also been done for baryons.  Furthermore, Poincar\'e-invariant form
factors can be obtained~\cite{Maris:2000sk,Bhagwat:2006pu} in a
limited momentum region without any ambiguity.  However, starting from
a purely Euclidean formulation, it is far from trivial to access
observables defined on the light-front, such as the the parton
distribution functions and their generalizations.

Here we remind the readers that with these continuum methods, it is
essential to take into account the nonperturbative dressing of quark
propagators and vertices, in particular for light mesons: the pions
represent the Goldstone bosons associated with dynamical chiral
symmetry breaking, and their Bethe--Salpeter amplitudes are closely
related to the self-energies of the light quarks~\cite{Maris:1997hd}.
Thus, if one aims to explore the rich kinematical range associated
with observable hadron structure, it is desirable to obtain
the solution of the BSE with dressed quark propagators in Minkowski
space.

To make progress with the DSEs applied to QCD, it is therefore
necessary to obtain the dressed propagators in Minkowski space.  The
DSE for the fermion self-energy within a QED-like model and
rainbow-ladder truncation has been studied extensively.  Early
investigations based on analytic continuation of the Euclidean DSE
suggested the existence of a pair of mass-like singularities at
complex-conjugate
momenta~\cite{Atkinson:1978tk,Maris:1993PhD,Maris:1994ux}.
Subsequently, the DSE was studied in Minkowski metric using the
Nakanishi integral representation (NIR)~\cite{NakaPTP69} in
Refs.~\cite{Sauli:2002tk,Sauli:2004bx,Sauli:2006ba}.  Their results
showed a complicated analytic structure of the self-energies in the
timelike region, which deserves to be studied further.  More recently,
the solutions for DSE for the fermion propagator in Minkowski space
with on-shell renormalization within quenched QED were obtained in
Ref.~\cite{Jia:2017niz}.

Efforts in solving the two-boson BSE in Minkowski space with bare
particles using the NIR have been undertaken since the pioneering
works in Refs.~\cite{Kusaka:1995za,Kusaka:1997xd}, which relied on the
uniqueness of the Nakanishi weight function in the nonperturbative
domain of bound states.  These techniques were further developed by
the introduction of the light-front projection allied to the NIR to
solve the BSE's for
bosons~\cite{Karmanov:2005nv,Sauli:2010ava,Frederico:2011ws,Frederico:2013vga}
and for fermions~\cite{Carbonell:2010zw,dePaula:2016oct,dePaula:2017ikc}.
Recently we obtained the approximate two-boson Minkowski
Bethe--Salpeter amplitude from the solution of the Euclidean BSE by
numerically `un-Wick rotating' the homogeneous integral equation
towards Minkowski space~\cite{Castro:2019tlh}.  The solutions found
with this new approach reveal the rich analytic structure of the
Bethe--Salpeter amplitude, consistent with the one obtained in
Minkowski space via the Nakanishi integral representation.

Motivated by the success of the un-Wick rotation method developed for
solving the BSE, and the challenge to obtain the self-energy in the
timelike region, this approach is extended here to investigate the
fermion self-energies both in the spacelike and the timelike regions.
We use the rainbow-ladder truncation of the fermion DSE with a massive
or massless exchange vector boson.  In Sec.~\ref{sec:DSE_MinkEucl},
the truncated DSE is presented with its representations both in
Minkowski metric and in Euclidean metric.  Here we restrict ourselves
to Feynman gauge, but the method is applicable in any covariant gauge.
We rely on Pauli--Villars (PV) regularization to eliminate ultraviolet
divergences; for simplicity we do not apply any renormalization
condition, so our numerical results depend on the PV mass.

We solve the truncated DSE in the complex momentum plane using two
different implementations:
\begin{enumerate}
\item the complex-rotation of the fourth component of the Euclidean
  four-momenta towards the zeroth component (energy component) of the
  four-momenta in Minkowski metric (`un-Wick rotation');
\item and an analytic continuation of the magnitude of the Euclidean
  four-momenta to rotate the Euclidean DSE on the spacelike axis
  towards the pure timelike axis in Minkowski metric,
\end{enumerate}
as described in Sec.~\ref{sec:solving_DSE}.  Both implementations give
(within their numerical uncertainty) the same results in a large
region of the complex momentum plane.  The numerical results for the
self-energies are discussed in Sect.~\ref{sec:results}.  In this
preliminary study, the coupling constant is chosen below the critical
value for dynamical chiral symmetry breaking, but large enough to
allow for nonperturbative effects.  We also demonstrate that the
obtained results close to the timelike axis can be used as a good
approximation to the spectral representation of the self-energy.

\section{DSE in Minkowski and Euclidean metric
  \label{sec:DSE_MinkEucl}}
In Minkowski metric we can write the inverse fermion propagator $S^{-1}$ as
\begin{eqnarray}
 S^{-1}(p) &=& p\!\!/ A(p^2) - B(p^2)
         \; = \; A(p^2) \, \left( p\!\!/ - M(p^2) \right)\, ,
\end{eqnarray}
with $M(p^2) = B(p^2)/A(p^2)$. For convenience we also define
$Z(p^2) = 1/A(p^2)$.  With this notation, the fermion propagator
$S$ can be written as
\begin{eqnarray}
  S(p) &=& \frac{A(p^2)\, p\!\!/ + B(p^2)}{A(p^2)\,p^2 - B^2(p^2) + i\epsilon}
      \;=\;  Z(p^2) \frac{ p\!\!/ + M(p^2)}{p^2 - M^2(p^2) + i\epsilon}\, ,
\end{eqnarray}
where we have introduced the $i\epsilon$ prescription to select the
correct Riemann sheet when the denominator in the spectral
representation vanishes.  For simplicity however, we will suppress the
explicit $i\epsilon$'s unless that could cause ambiguities.

Next, consider DSE for the fermion propagator in the rainbow (ladder)
truncation by coupling to a vector boson with mass $\mu$ and
PV regularization with mass $\Lambda$
\begin{eqnarray}
  S^{-1}(p) &=& p\!\!/ - m_0
  - i g^2 \int \frac{d^4k}{(2\pi)^4} \gamma^\mu \, S(k) \, \gamma^\nu \,
  \left[D_{\mu\nu}(q;\mu) - D_{\mu\nu}(q;\Lambda)\right]\, ,
\end{eqnarray}
with the bare fermion mass $m_0$ and $q = p - k$.  The (massive)
vector boson in the covariant gauge can be written as~\cite{IZ}
\begin{eqnarray}
  D_{\mu\nu}(q;m) &=& \frac{-1}{q^2 - m^2+i\epsilon} \;
  \left[g_{\mu\nu} - (1-\xi) \frac{q_\mu q_\nu}{q^2-\xi\,m^2+i\epsilon}\right]\, ,
\end{eqnarray}
where $\xi$ is the gauge parameter.  The Landau gauge is defined by
$\xi = 0$, while $\xi=1$ defines the Feynman gauge.  For simplicity,
we will only consider Feynman gauge here.  Projecting out the
equations for $A$ and $B$ we arrive at
\begin{eqnarray}
  B(p^2) &=& m_0 + i g^2 \int \frac{d^4k}{(2\pi)^4} \,
  \frac{4\,B(k^2)}{k^2 \, A^2(k^2) - B^2(k^2)} \, 
   \frac{\Lambda^2-\mu^2}{(q^2 - \mu^2)(q^2 - \Lambda^2)} \, ,
  \label{DSE_B_Feynman}
  \\
  A(p^2) &=& 1 + i g^2 \int \frac{d^4k}{(2\pi)^4} \,
  \frac{2\, p\cdot k}{p^2}  \,
  \frac{A(k^2)}{k^2 \, A^2(k^2) - B^2(k^2)} \, 
  \frac{\Lambda^2-\mu^2}{(q^2 - \mu^2)(q^2 - \Lambda^2)} \, ,
  \label{DSE_A_Feynman}
\end{eqnarray}
with implicit $i\epsilon$ prescriptions for the various propagator
poles.

Solving the DSE numerically directly in Minkowski space poses the
following challenges:
\begin{itemize}
\item the integration $\int d^4k$ in Minkowski metric;
\item the known singularities in the denominators $(q^2 - \mu^2)$ and $(q^2 - \Lambda^2)$;
\item the unknown but expected singularity in the denominator $k^2 \, A^2(k^2) - B^2(k^2)$.
\end{itemize}
The first challenge can be dealt with by integrating over $k_0$ and
$\vec{k}$ separately:
\begin{eqnarray}
\int \frac{d^4k}{(2\pi)^4} &=&
\int_{-\infty}^\infty\!\frac{dk_0}{2\pi} \; \int \frac{d^3\vec{k}}{(2\pi)^3}\, .
\end{eqnarray}
The latter two could be overcome by using an explicitly nonzero
$i\,\epsilon$ in the propagators denominators.  However, numerically
this is not necessarily stable, in particular since the location of
the singularity in the fermion propagator is determined by the
solution of the DSE.

Indeed, the common practice is to perform a formal Wick rotation to
Euclidean space, avoiding the singularities alltogether.  Of course,
the DSE can only be solved for Euclidean momenta after such a
procedure, corresponding to spacelike momenta in Minkowski metric.
Specifically, after applying the formal Wick rotation, we obtain the
fermion DSE using Euclidean four-vectors $p{}_{\mathrm{E}}$ and
$k{}_{\mathrm{E}}$
\begin{eqnarray}
&& B(-p_{\mathrm{E}}^2) = m_0 + g^2 \!\int\! \frac{d^4k_{\mathrm{E}}}{(2\pi)^4} \,
  \frac{4\,B(-k_{\mathrm{E}}^2)}{k_{\mathrm{E}}^2 \, A^2(-k_{\mathrm{E}}^2) + B^2(-k_{\mathrm{E}}^2)} \, 
  \frac{\Lambda^2-\mu^2}{(q_{\mathrm{E}}^2 + \mu^2)(q_{\mathrm{E}}^2 + \Lambda^2)} \, ,
  \\
 && A(-p_{\mathrm{E}}^2) = 1 + g^2 \!\int\! \frac{d^4k_{\mathrm{E}}}{(2\pi)^4} \,
  \frac{A(-k_{\mathrm{E}}^2)}{k_{\mathrm{E}}^2 \, A^2(-k_{\mathrm{E}}^2) + B^2(-k_{\mathrm{E}}^2)} \, 
  \frac{2\, p_{\mathrm{E}}\cdot k_{\mathrm{E}} \; (\Lambda^2-\mu^2)}
       {p_{\mathrm{E}}^2\,(q_{\mathrm{E}}^2 + \mu^2)(q_{\mathrm{E}}^2 + \Lambda^2)} \, .
\end{eqnarray}
Note that in Euclidean metric, $p_{\mathrm{E}}^2$ runs from $0$ to
$+\infty$, and that results for Euclidean $p_{\mathrm{E}}^2 \ge 0$ are
equivalent to results for spacelike momenta $p^2 = - p_{\mathrm{E}}^2 \le 0$
in Minkowski metric.  In the next section we discuss how one can
obtain the solution of the DSE for timelike momenta.

\section{Solving the DSE numerically
  \label{sec:solving_DSE}}
In Euclidean space we can perform the integrations using 4-dimensional
hyperspherical coordinates:
\begin{eqnarray}
  \int \frac{d^4k_{\mathrm{E}}}{(2\pi)^4} &=&
  \int_0^\infty \frac{k^3_{\mathrm{E}} \, dk_{\mathrm{E}}}{(2\pi)^4}
  \int_0^\pi \sin^2(\theta)\, d\theta \;
  \int_0^\pi \sin(\phi)\, d\phi \;
  \int_0^{2\pi}  d\alpha \; .
\end{eqnarray}
The unknown functions $A$ and $B$ of the fermion propagator only
depend on $k^2$, and there is only one nontrivial angle in the
integrand, namely the angle between $k$ and $p$.  Thus we can perform
two of the three angular integrations analytically, with the remaining
angular integral to be evaluated numerically
\begin{eqnarray}
  \int \frac{d^4k_{\mathrm{E}}}{(2\pi)^4} \; I(k, p) &=&
  2 \int_0^\infty \frac{k^3_{\mathrm{E}} \, dk_{\mathrm{E}}}{(2\pi)^3}
  \int_0^\pi \sin^2(\theta)\, d\theta \; I(k^2_{\mathrm{E}}, p^2_{\mathrm{E}}, \cos(\theta))
\end{eqnarray}
This leads to a set of coupled nonlinear integral equations in one
dimension for spacelike values of $p^2_{\mathrm{E}} \ge 0$
\begin{eqnarray}
\label{dsefgsl}
  B(-p^2_{\mathrm{E}}) &=& m_0 + \frac{2\,g^2}{(2\pi)^3}  \int_0^\infty k^3_{\mathrm{E}} \, dk_{\mathrm{E}}
  \frac{4\,B(-k^2_{\mathrm{E}})}{k^2_{\mathrm{E}} \, A^2(-k^2_{\mathrm{E}}) + B^2(-k^2_{\mathrm{E}})} \;
  \nonumber \\ && {} \times \int_0^\pi \sin^2\theta\, d\theta \; 
  \frac{\Lambda^2-\mu^2}{(q^2_{\mathrm{E}} + \mu^2 ) (q^2_{\mathrm{E}} + \Lambda^2)} \, ,
  \\
  A(-p^2_{\mathrm{E}}) &=& 1 + \frac{2\,g^2}{(2\pi)^3} \int_0^\infty k^3_{\mathrm{E}} \, dk_{\mathrm{E}}
  \frac{A(-k^2_{\mathrm{E}})}{k^2_{\mathrm{E}} \, A^2(-k^2_{\mathrm{E}}) + B^2(-k^2_{\mathrm{E}})} \, 
  \nonumber \\ && {} \times\int_0^\pi \sin^2\theta\, d\theta \,
  \frac{2 \, k_{\mathrm{E}} \, \cos\theta}{p_{\mathrm{E}}} \,
  \frac{\Lambda^2-\mu^2}{(q^2_{\mathrm{E}} + \mu^2 ) (q^2_{\mathrm{E}} + \Lambda^2)} \, .
\end{eqnarray}
It is straightforward to solve these coupled nonlinear integral
equations iteratively using a suitable discretization of the integrals
and an initial guess for the functions $A$ and $B$.

\subsection{Un-Wick rotating from the Euclidean solution
  \label{sec:unWick}}
Instead of using 4-dimensional hyperspherical coordinates, we can also integrate
over the fourth (or energy) component separately, and use
3-dimensional spherical coordinates for the remaining 3 dimensions
\begin{eqnarray}
  \int \frac{d^4k_{\mathrm{E}}}{(2\pi)^4} &=&
  \int_{-\infty}^\infty\!\frac{dk_4}{2\pi} \; \int \frac{d^3\vec{k}}{(2\pi)^3}
  \\
  &=& \frac{1}{(2\pi)^3}
  \int_{-\infty}^\infty\! dk_4 \; \int_0^{\infty}\! k_v^2 dkv \, \int_0^\pi \sin(\phi)\, d\phi\, ,
\end{eqnarray}
where $k_v = |\vec{k}|$.  In this case, it is convenient to write the
inverse of the fermion propagator $A$ and $B$ as functions of two
variables, $p_4$ and $p_v$.  After doing so, we arrive at
\begin{eqnarray}
  B(p_4, p_v) &=& m_0 + \frac{g^2}{(2\pi)^3}\int_{-\infty}^\infty\! dk_4 \int_0^{\infty}\! k_v^2 dkv \, 
               \frac{4\,B(k_4, k_v)   }{(k_4^2 + k_v^2) \, A^2(k_4, k_v) + B^2(k_4, k_v)} \; 
               \nonumber \\  && {}
         \times  \int_0^\pi \sin(\phi)\, d\phi \,
         \frac{\Lambda^2-\mu^2}{(q^2_{\mathrm{E}} + \mu^2 ) (q^2_{\mathrm{E}} + \Lambda^2)} \, ,
  \label{DSE_B_p0pv}
         \\
  A(p_4, p_v) &= & 1 + \frac{g^2}{(2\pi)^3} \int_{-\infty}^\infty\! dk_0 \int_0^{\infty}\! k_v^2 dkv \, 
               \frac{A(k_4, k_v)}{(k_4^2 + k_v^2) \, A^2(k_4, k_v) + B^2(k_4, k_v)} \;
               \nonumber \\ && {}
              \times \int_0^\pi \sin(\phi)\, d\phi \,
               \frac{2\,(p_4k_4 + p_v k_v\cos\phi)}{p_4^2+p_v^2} \,
               \frac{\Lambda^2-\mu^2}{(q^2_{\mathrm{E}} + \mu^2 ) (q^2_{\mathrm{E}} + \Lambda^2)}  \, ,
  \label{DSE_A_p0pv}
\end{eqnarray}
where
$q^2_{\mathrm{E}} = (p_4-k_4)^2 + (\vec{p}-\vec{k})^2 = p_4^2 - 2p_4k_4 + k_4^2 + p_v^2 - 2p_vk_v\cos(\phi) + k_v^2$.
We can now solve for $A$ and $B$ as functions of two variables
$p_4$ and $p_v$, and up to numerical precision, we should get the
same results for $A(p_4^2 + p_v^2)$ and $B(p_4^2 + p_v^2)$ as above.

We can now undo the Wick rotation by applying the transformation
\begin{eqnarray}
  p_4 \to {\rm e}^{-i\delta} \, p_4\, ,\quad
  k_4 \to {\rm e}^{-i\delta} \, k_4
 \,,\quad
  dk_4 \to {\rm e}^{-i\delta} \, dk_4\, ,
  \label{Eq:unWick}
\end{eqnarray}
while keeping $p_4$ and $k_4$ real, analogous to the method used in
Ref.~\cite{Castro:2019tlh} to obtain the Minkowski space
Bethe--Salpeter amplitudes from the Euclidean BSE.  As long as the
contribution from the integral along the arcs at $\vert k_4\vert
=\pm\infty$ vanishes, true in the case of PV regularization, we only
need to keep the integration over $k_4$ from $-\infty$ to $\infty$.

In the limit of $\delta \to \pi/2$ this transformation becomes
\begin{eqnarray}
  p_4 \to -i \, p_4\, \equiv p_0 \, ,\quad
  k_4 \to -i \, k_4\, \equiv k_0 \, ,\quad
  dk_4 \to -i \, dk_4\, \equiv dk_0 \, ,
\end{eqnarray}
which recovers the DSEs in Minkowski metric, for both
the spacelike and the timelike region.  Indeed, applying this
transformation to Eqs.~(\ref{DSE_B_p0pv}) and (\ref{DSE_A_p0pv})
we obtain
\begin{eqnarray}
  B(p_0, p_v) &=& m_0 + i \frac{g^2}{(2\pi)^3} \! \! \int_{-\infty}^\infty\! \! dk_0 \int_0^{\infty}\! \! \! k_v^2 dkv \, 
               \frac{4\,B(k_0, k_v)}{(k_0^2 - k_v^2) \, A^2(k_0, k_v) - B^2(k_0, k_v)} \; 
   \nonumber \\ && {}  
   \times              \int_0^\pi \sin(\phi)\, d\phi
             \frac{\Lambda^2-\mu^2}{(-q_0^2 + q_v^2 + \mu^2 ) (-q_0^2 + q_v^2 + \Lambda^2)}  \, ,
               \\
  A(p_0, p_v) &=& 1 + i \frac{g^2}{(2\pi)^3} \int_{-\infty}^\infty\! dk_0 \int_0^{\infty}\! k_v^2 dkv \, 
               \frac{A(k_0, k_v)}{(k_0^2 - k_v^2) \, A^2(k_0, k_v) - B^2(k_0, k_v)} \;
               \nonumber \\ && {}
               \times \int_0^\pi \sin(\phi)\, d\phi
               \frac{p_0k_0 - p_vk_v\cos\phi}{p_0^2 - p_v^2} \,
              \frac{\Lambda^2-\mu^2}{(q_0^2 - q_v^2 - \mu^2 ) (q_0^2 - q_v^2 - \Lambda^2)}  \, ,
   \nonumber \\ && {}  
\end{eqnarray}
where $q_0^2 = (p_0-k_0)^2$ and $q_v^2 = (\vec{p}-\vec{k})^2$.  Now we
can recognize $p_0^2 - p_v^2$ as $p^2$ in Minkowski metric, and
similarly for $k_0^2 - k_v^2$ and $q_0^2 - q_v^2$, and thus we arrive
at the DSE in Minkowski space, Eqs.~(\ref{DSE_B_Feynman}) and
(\ref{DSE_A_Feynman}).  Of course, in these expressions for the DSEs
in Minkowski metric for both timelike and spacelike momenta, there are
singularities in the propagators under the integral, which are
understood in conjunction with $i\epsilon$ prescription.

With $\delta \in (0, \pi/2)$, the transformation given by
Eq.~(\ref{Eq:unWick}) acts as the tool to interpolate the DSEs
between Euclidean and Minkowski metrics.
In the limit of $\delta \to \pi/2$, the Minkowski space invariant
$p^2 = p_0^2 - p_v^2$ is real, and runs from $-\infty$ to $+\infty$.
But for $0 < \delta < \pi/2$ the `invariant'
$p^2 = -{\rm e}^{-2i\delta} \, p_4^2 - p_v^2$ covers a slice in the
upper complex $p^2$ plane.  As $\delta$ approaches $\pi/2$, it covers
almost the entire upper complex momentum plane, and `collapses' onto
the real axis only in the limit $\delta \to \pi/2$.  As long as
there are no singularities in the upper complex $p^2$ plane, we can
continuously connect the solution of the DSEs near the timelike region
to the solution in the spacelike region.  As a consistency check,
for any value of $0 \ge \delta \ge \pi/2$ we should obtain the same
(spacelike) solution for $p_4=0$.

\begin{figure}
  \includegraphics[width=0.48\columnwidth]{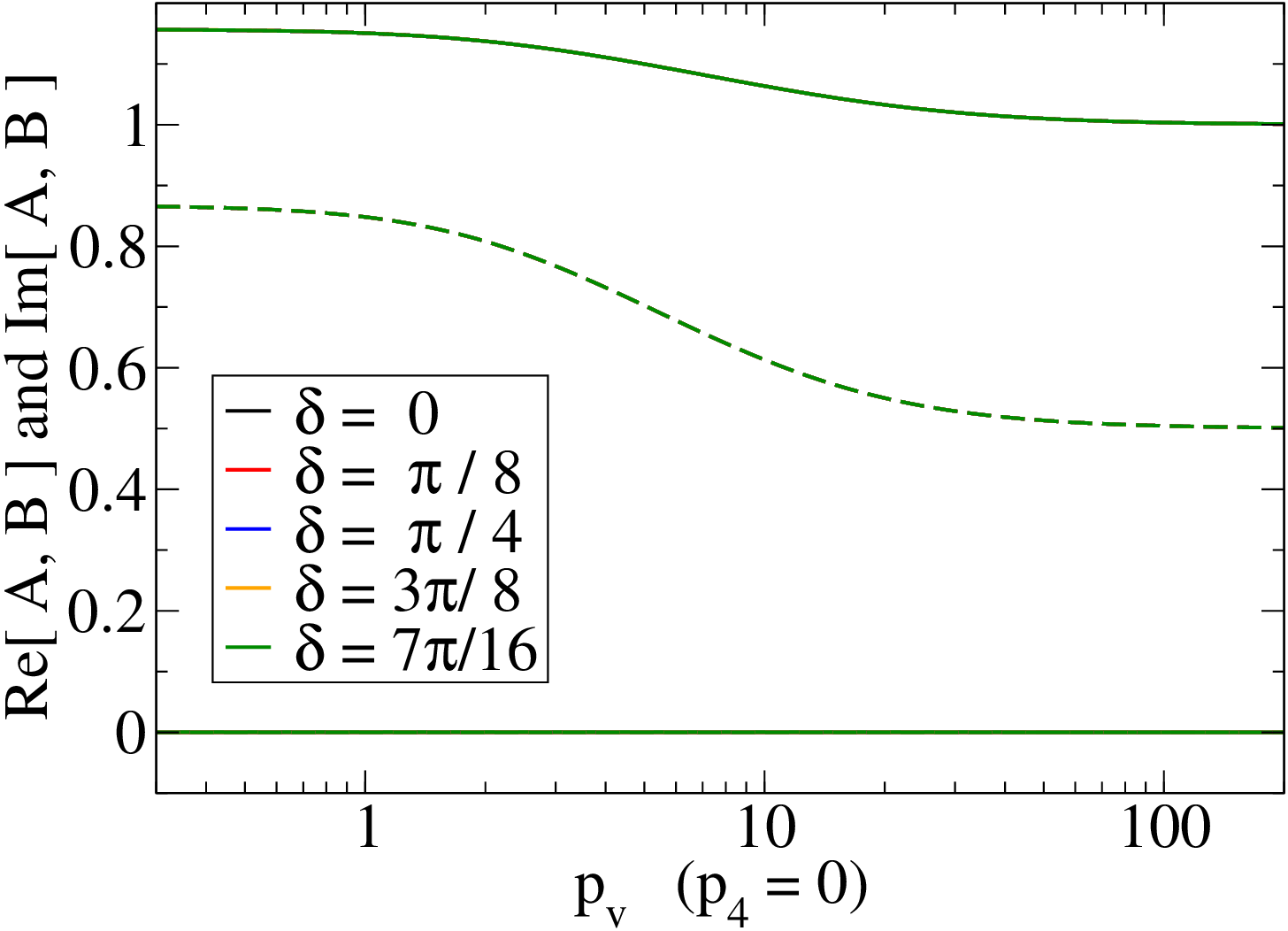} \quad
  \includegraphics[width=0.48\columnwidth]{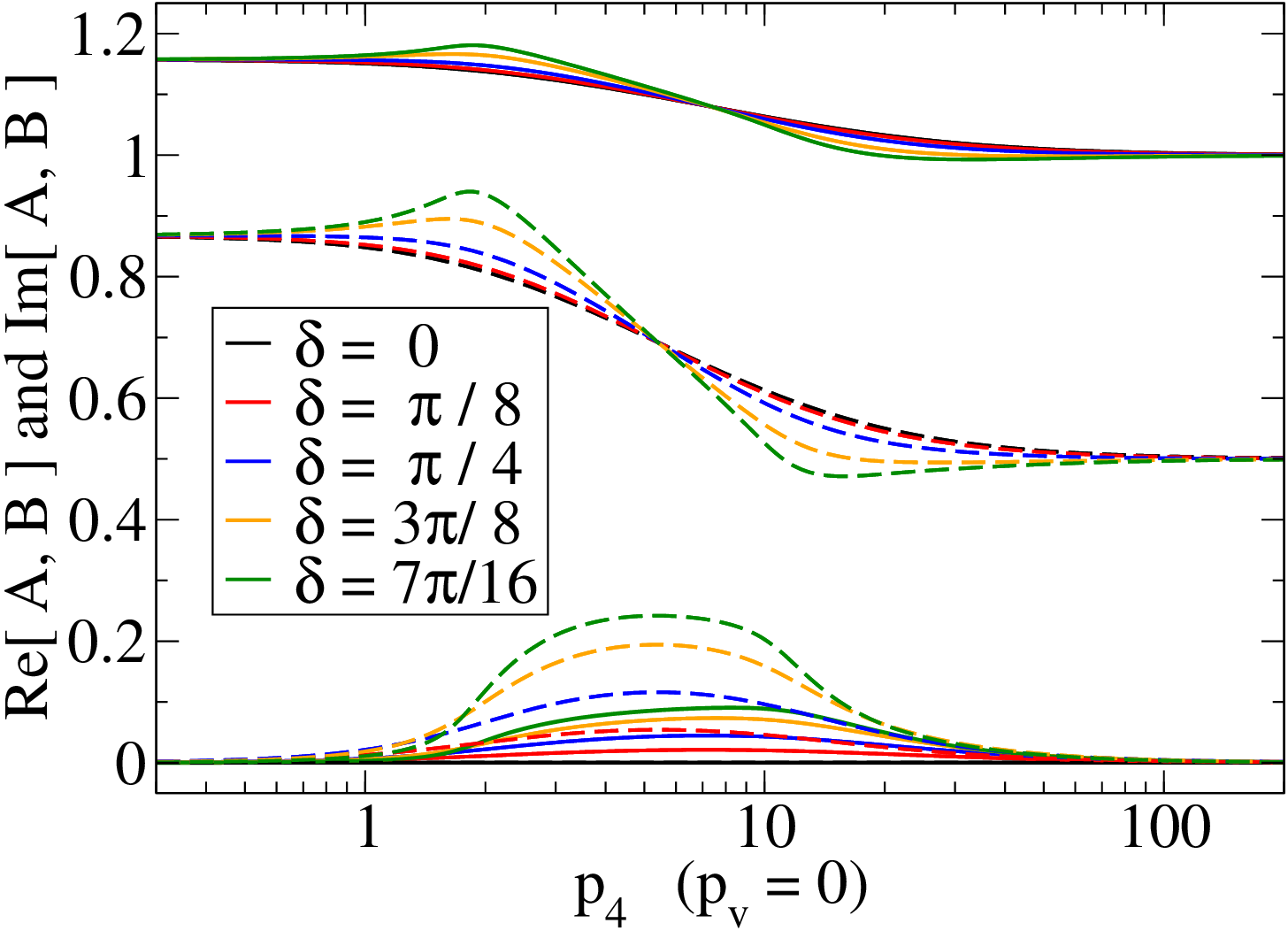}
  \caption{ \label{Fig:AB_space_time}
    Real and imaginary parts of the inverse propagator functions
    $A$ (solid) and $B$ (dashed), at different angles $\delta$,
    obtained by un-Wick rotating the Euclidean solution as function
    of $p_v$ at $p_4 = 0$, corresponding to the spacelike $p^2$ axis
    (left) and as function of $p_4$ at $p_v = 0$, along a line in the
    complex $p_4 {\rm e}^{-i\,\delta}$ plane (right); $\delta = \pi/2$
    would be the timelike axis.  On the right we also show our results
    of rotating the magnitude of $p$ from the spacelike region towards
    the timelike region, which are indistinguishable at the scale
    shown.  Parameters are $m_0 = 0.5$, $\mu=1.0$, $\Lambda = 10.0$,
    and $\alpha = 0.5$.  }
\end{figure}
In Fig.~\ref{Fig:AB_space_time} we present solutions of the DSE in the
Feynman gauge obtained by un-Wick rotating $p_4$.  When un-Wick
rotating $p_4$ from Euclidean metric, we solve the DSE on a slice in
the complex $p^2= {\rm e}^{i\,2\delta} p_4^2 + p_v^2$ plane; the
boundaries of this slice are given by $(p_4=0, p_v)$, which
corresponds to the spacelike axis, and by $(p_4, p_v=0)$, which
approaches the timelike axis in the limit $\delta \to \pi/2$.  The
results for $A(p_4=0, p_v)$ and $B(p_4=0, p_v)$, i.e.  on the
spacelike axis, are indeed independent of the angle $\delta$, and
purely real, as is shown in the left panel of
Fig.~\ref{Fig:AB_space_time}.  In the right panel, we show our results
as function of $p_4$ for $p_v=0$, in wich case we do see a dependence
on the angle $\delta$, as expected; furthermore, both $A$ and $B$
develop an imaginary part, which increases in magnitude with
increasing $\delta$.  However, as we approach $\delta = \pi/2$,
the numerics becomes unstable due to singularities in the propagators,
which prevents us from actually reaching the timelike axis.

\subsection{Rotating the spacelike region to the timelike region
  \label{sec:space2time}}
Alternatively, we can rotate the DSE from the Euclidean spacelike axis
towards the timelike axis by applying the transformation
\begin{eqnarray}
\label{psltl}
p \to {\rm e}^{-i\delta} \, p\, ,\quad
  k \to {\rm e}^{-i\delta} \, k\, ,\quad
  dk \to {\rm e}^{-i\delta} \, dk\, ,
\end{eqnarray}
on the magnitude of the (Euclidean) four-vectors, while continuing to
use 4-dimensional hyperspherical coordinates, as was done in
e.g. Refs.~\cite{Maris:1993PhD,Maris:1994ux}.  With this technique we
keep $p$ and $k$ real (and positive), and we retain the 4-dimensional
symmetry.  As long as the contribution along the arc at $k=\infty$
vanishes (and with the explicit PV regularization it does), we can
neglect the contribution along this arc, and keep only the integration
over $k$ from $0$ to $\infty$.

In the limit of $\delta = \pi/2$ this transformation becomes
\begin{eqnarray}
  p^2_{\mathrm{E}} \to -p^2_{\mathrm{E}} = p^2 \, , \quad
  k^2_{\mathrm{E}} \to -k^2_{\mathrm{E}} = k^2 \, \quad
  k_{\mathrm{E}}^3\, dk_{\mathrm{E}} \to k_{\mathrm{E}}^3\, dk_{\mathrm{E}} = k^3 \, dk\, ,
\end{eqnarray}
and effectively this gives us a the DSEs on the pure timelike axis with
$p^2 \ge 0$
\begin{eqnarray}
  B(p^2) &=& m_0 - \frac{2\,g^2}{(2\pi)^3} \int_0^\infty k^3 \, dk
  \frac{4\,B(k^2)}{k^2 \, A^2(k^2) - B^2(k^2)} \;
  \nonumber  \\ && \times 
               \int_0^\pi \sin^2\theta\, d\theta \; 
               \frac{\Lambda^2-\mu^2}{(q^2 - \mu^2 ) (q^2 - \Lambda^2)}  \, ,
                \label{Eq:timelikeB}
               \\
  A(p^2) &=& 1 - \frac{2\,g^2}{(2\pi)^3} \int_0^\infty k^3 \, dk
               \frac{A(k^2)}{k^2 \, A^2(k^2) - B^2(k^2)} \,
  \nonumber  \\ && \times 
                \int_0^\pi \sin^2\theta\, d\theta \,
               \frac{2\,k\cos\theta}{p} \,
               \frac{\Lambda^2-\mu^2}{(q^2 - \mu^2 ) (q^2 - \Lambda^2)}  \, .
               \label{Eq:timelikeA}
\end{eqnarray}
Note Eqs.~(\ref{Eq:timelikeB}) and (\ref{Eq:timelikeA}) are for
timelike momenta only, $p^2 \ge 0$, $k^2 \ge 0$, and $q^2=(p-k)^2 \ge 0$
 -- they are different from the DSEs in Minkowski metric,
Eqs.~(\ref{DSE_B_Feynman}) and (\ref{DSE_A_Feynman}).  (Again
singularities under the integrals are specified by the $i\epsilon$
prescription.)

For any $0 < \delta < \pi/2$, this method gives the DSE along a line
from $0$ to $\infty$ in the upper complex $p^2$ plane, rather than on
a slice of the upper complex momentum plane.  Furthermore, it remains
an integral equation in one variable, rather than in two variables as
with the method described in the previous subsection.  This method is
therefore numerically easier to implement, and leads to better
numerical precision.

In the right panel of Fig.~\ref{Fig:AB_space_time}, we also include
our results obtained with this method.  Not surprisingly, the results
of the two methods are essentially indistinguishable, at least at the
scale shown.  However, the method of rotating the magnitude of $p$ is
much more accurate (for a similar numerical effort) than the explicit
un-Wick rotation of the fourth component, because when we un-Wick
rotate the fourth component, we break the 4-dimensional symmetry by
treating the fourth component and the 3-vector components differently.
Furthermore, we solve the propagator functions $A$ and $B$ as
functions of two independent real variables, $p_4$ and $p_v$,
for a given angle $\delta$ (or equivalently, as function of one
complex variable $p^2= p_4^2 {\rm e}^{i\,2\delta} + p_v^2$),
whereas if we rotate the magnitude of $p$ the functions $A$ and $B$
remain function of only one essentially real variable.  In particular,
as $\delta$ approaches $\pi/2$, in the case of the un-Wick rotation we
solve the DSE in the entire upper $p^2$ plane, whereas if we rotate
the magnitude of $p$, we solve the DSE along a line from $0$ to
$\infty$ close to the timelike axis.  Clearly, the latter approach is
more efficient numerically.  

\section{Results for the self-energy in the timelike region
  \label{sec:results}}
\begin{figure}[bth]
  \includegraphics[width=0.48\columnwidth]{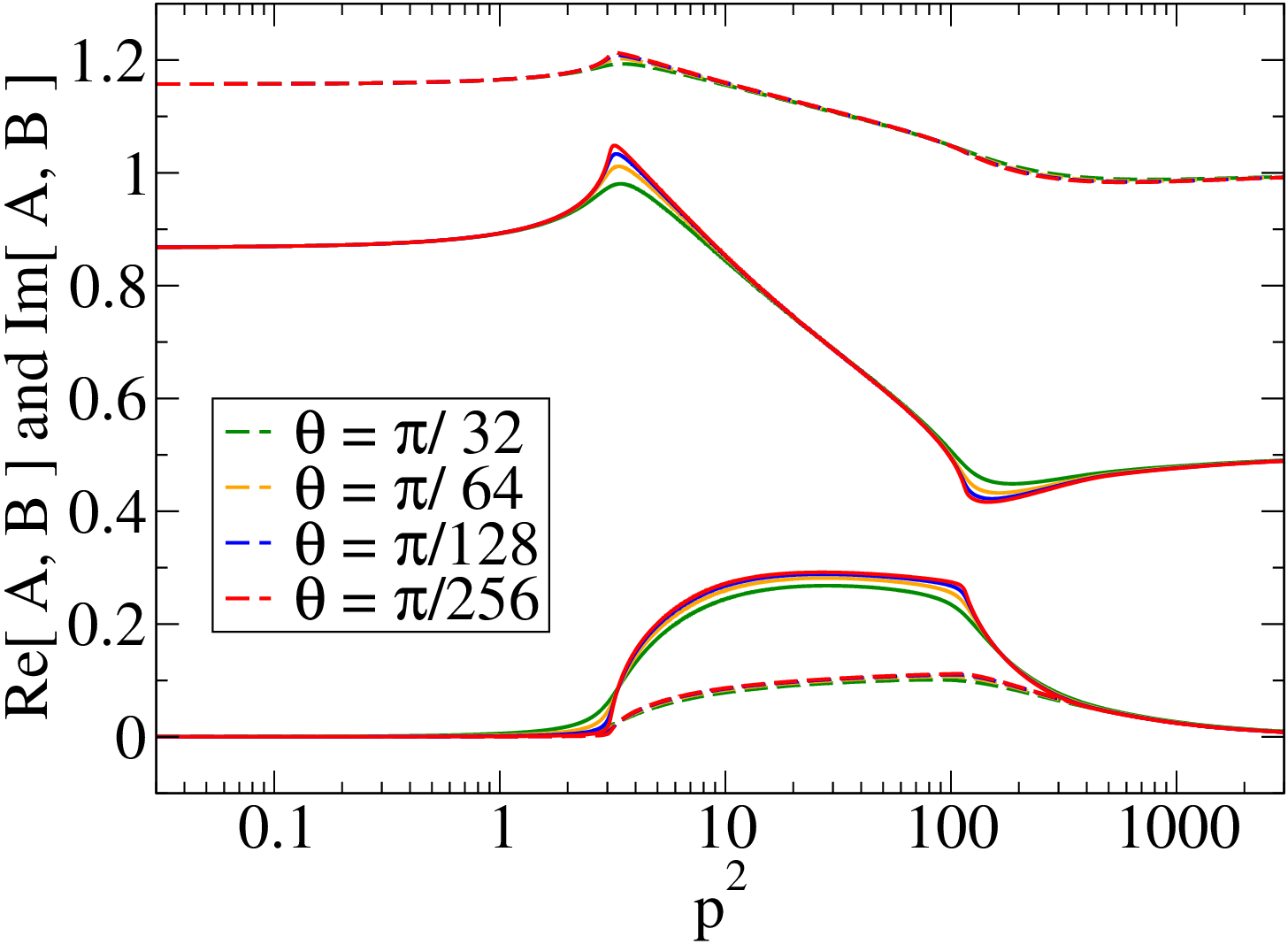}  \quad
  \includegraphics[width=0.48\columnwidth]{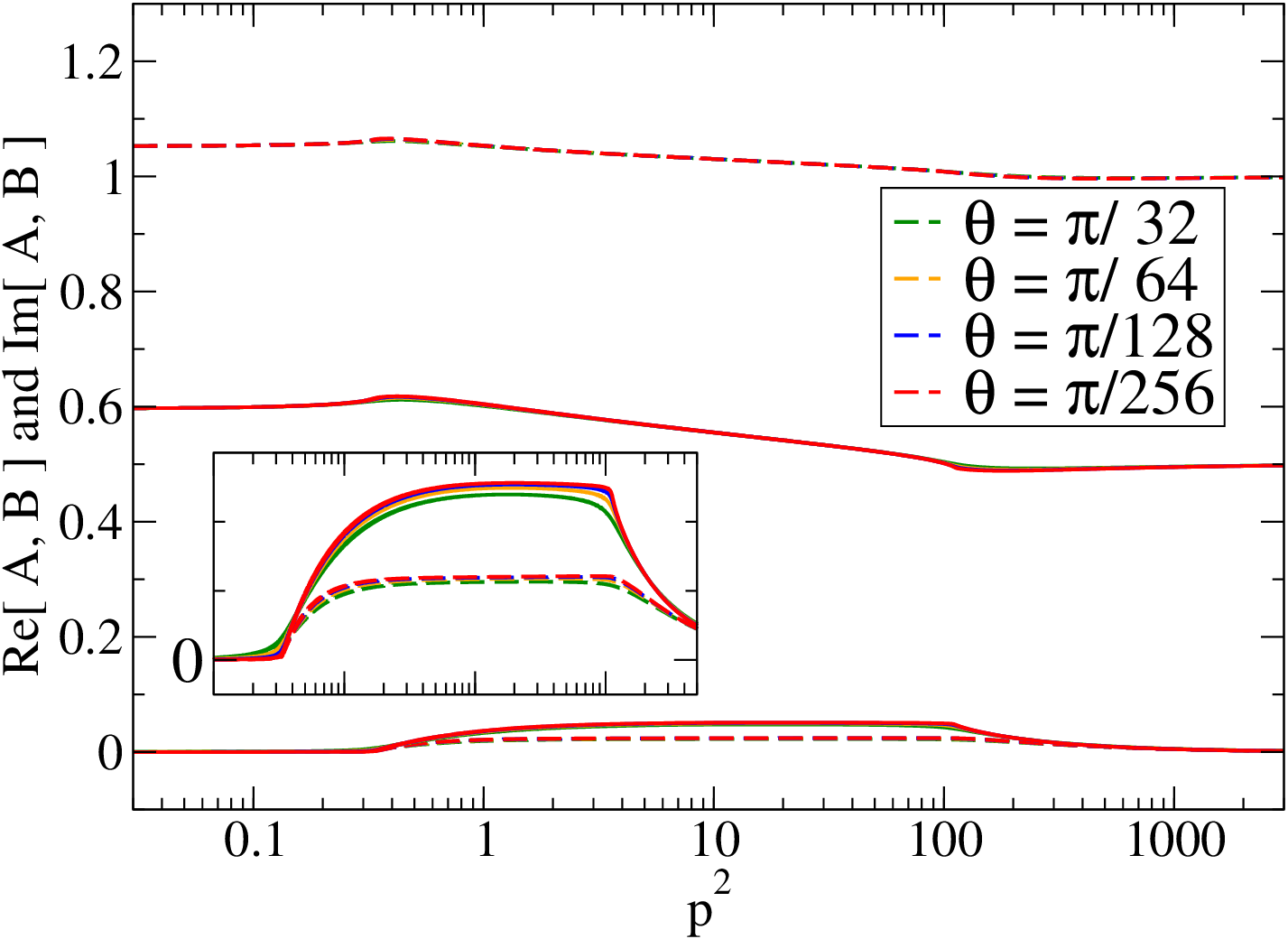} 
  \caption{ \label{Fig:AB_almost_timelike}
    Real and imaginary parts of the inverse propagator functions
    $A$ (dashed) and $B$ (solid), at different angles $\theta$
    close to the timelike axis.
    Both figures are with $m_0 = 0.5$ and PV mass $\Lambda = 10$;
    and exchange mass $\mu = 1.0$ and $\alpha = 0.5$ (left)
    and massless vector boson and $\alpha = 0.1$ (right).
  }    
\end{figure}
In order to discuss our results as we approach the timelike region,
it is more convenient to use $\theta=\pi/2-\delta$; with this notation
the timelike axis corresponds to the limit $\theta \to 0$.
For moderate values of the coupling (well below those corresponding
to dynamical chiral symmetry breaking), we can achieve accurate
results down to $\theta = \pi/256 \approx 0.7^{\circ}$ by rotating
the magnitude of $p$, whereas if we decrease $\theta$ below about
$\theta = \pi/16 \approx 11^{\circ}$, the un-Wick rotation becomes
numerically challenging, requiring an effecient implementation on
parallel high-performance computing systems.

In Fig.~\ref{Fig:AB_almost_timelike} we see that the imaginary parts
of $A(p^2)$ and $B(p^2)$ become nonzero along the timelike axis.
Furthermore, both the real parts and the imaginary parts of $A(p^2)$
and $B(p^2)$ develop kinks, that is, discontinuities in their
derivatives.  The location of these kinks is determined by the
physical thresholds for the production of an exchange particle;
these kinks occur at $(m_{\hbox{\scriptsize phys}} + \mu)^2$ and
$(m_{\hbox{\scriptsize phys}} + \Lambda)^2$, where the pole mass
$m_{\hbox{\scriptsize phys}}$ is determined from the zero of the
inverse propagator, at $M(p^2) = \sqrt{p^2}$.  

These kinks are generally attributed to the integration over the
propagator poles in Eqs.~(\ref{DSE_B_Feynman}) and
(\ref{DSE_A_Feynman}), where one (or more) denominator becomes zero.
Mathematically, the kinks are caused by a pinch singularity due to the
zeros of the exchange boson propagator and the fermion propagator in
Eqs.~(\ref{DSE_B_Feynman}) and (\ref{DSE_A_Feynman}).   

\subsection{Analytic structure and pole mass
  \label{sec:poles}}
\begin{figure}
  \includegraphics[width=0.48\columnwidth]{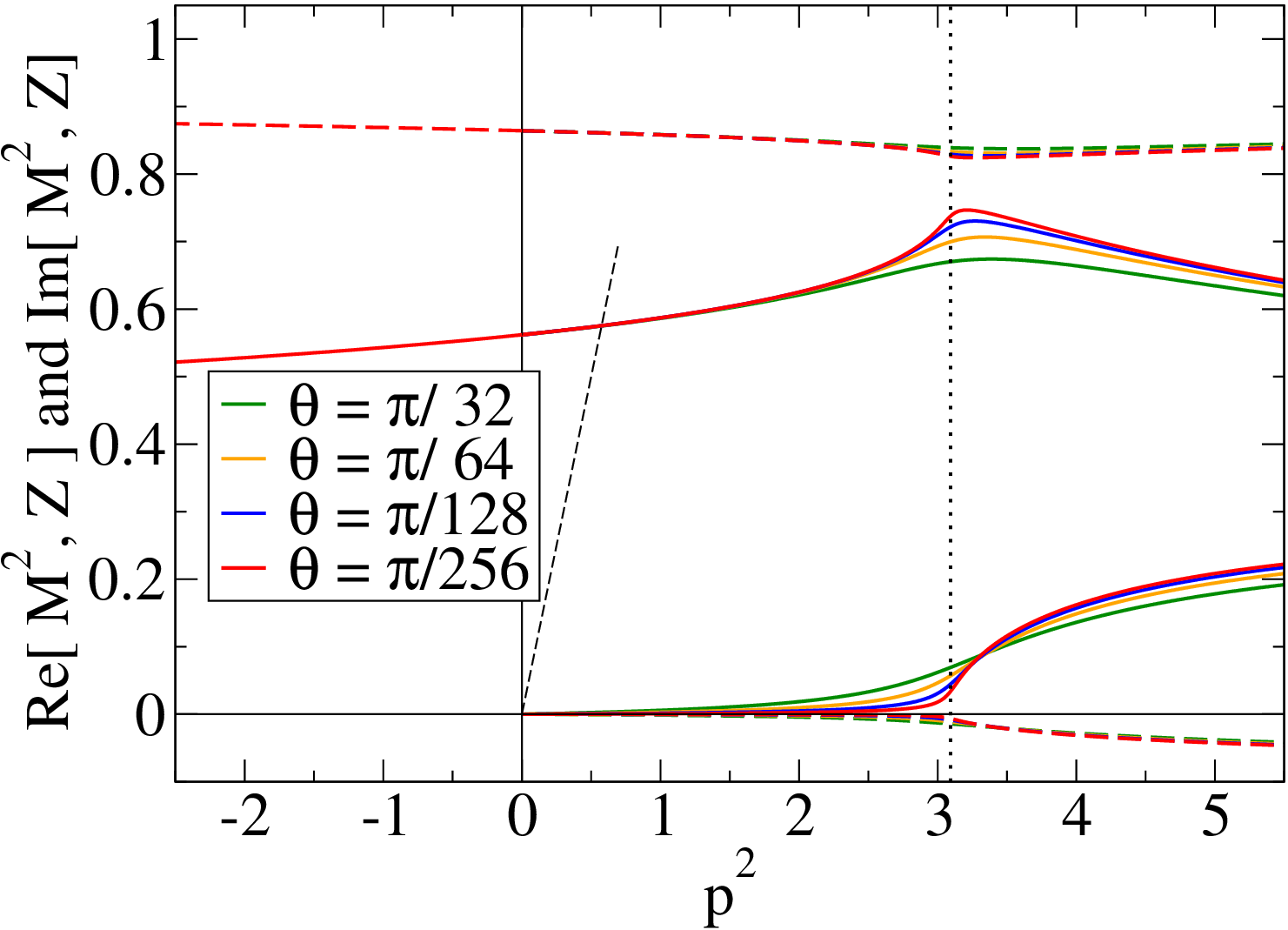}  \quad
  \includegraphics[width=0.48\columnwidth]{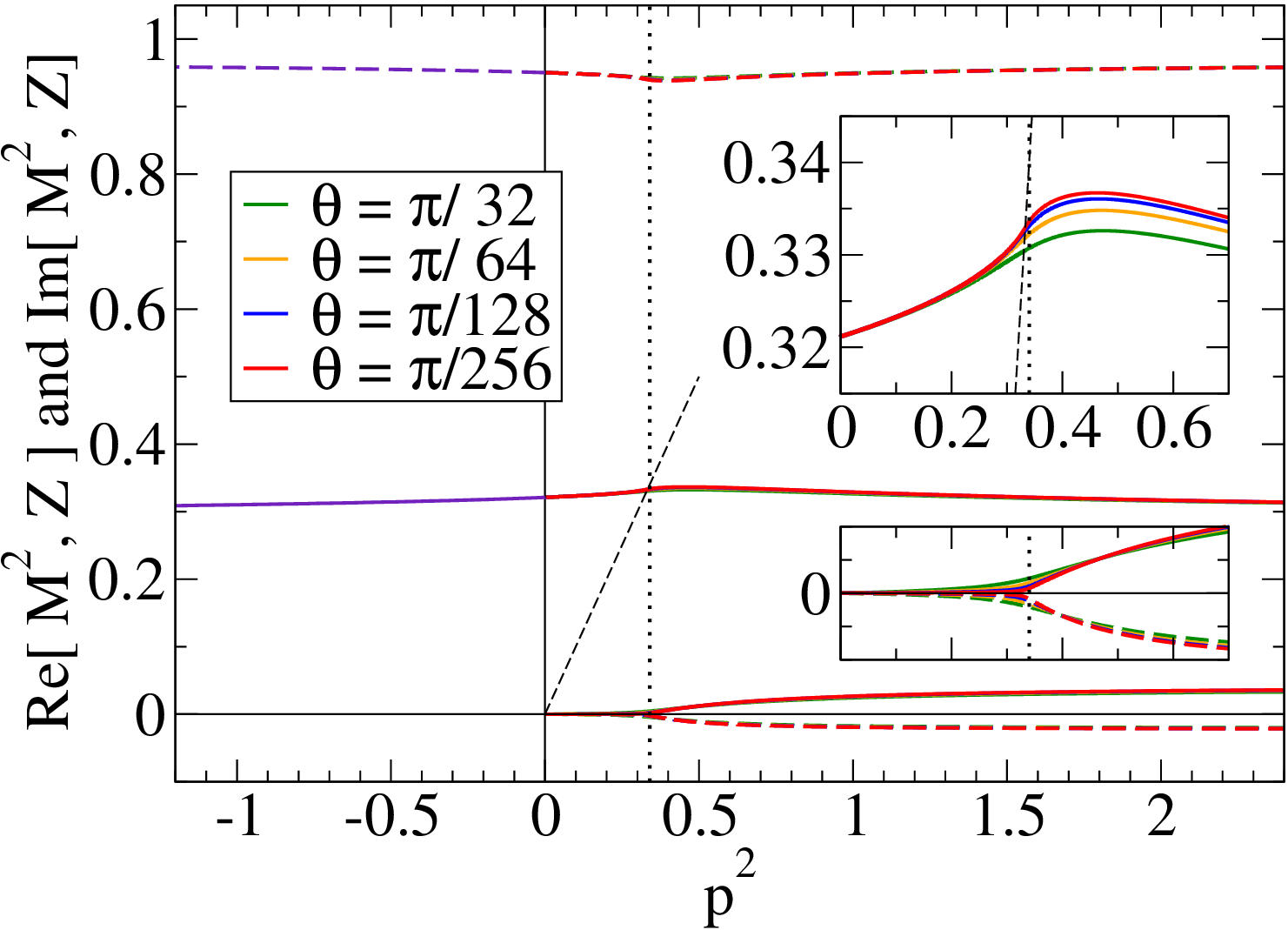} 
  \caption{ \label{Fig:MsqZ_almost_timelike}
    Real and imaginary parts of the dynamical mass squared, $M^2(p^2)$ (solid),
    and wave function renormalization, $Z(p^2)=1/A(p^2)$ (dashed),
    in the spacelike and close to the timelike axis,
    again with $m_0 = 0.5$ and PV mass $\Lambda = 10$.
    The extracted pole masses and residues are:
    $m=0.759$ and $Z(m^2)= 0.82$ for mass $\mu = 1.0$ and $\alpha = 0.5$ (left)
    and $m=0.58$ and $Z(m^2)= 0.34$ for massless vector boson and $\alpha = 0.1$ (right).
  }
\end{figure}
In Fig.~\ref{Fig:MsqZ_almost_timelike} we show our results for
$M^2(p^2)$ and $Z(p^2)=1/A(p^2)$ in the infrared region.  The fermion
propagator has a singularity at
$p^2 = M^2(p^2) = m^2_{\hbox{\scriptsize phys}}$ in the timelike
region.  With a nonzero mass for the exchange boson, this singularity
is a simple mass-pole (at least in Feynman gauge) -- but neither the
inverse propagator functions $A^2(p^2)$ and $B(p^2)$, nor the
dynamical mass function $M(p^2)$ shows any discontinuity or kink at
this mass-pole.

The first kink or branch-point in the inverse propagator functions is
located at
$(m_{\hbox{\scriptsize phys}} + \mu)^2 \ge m^2_{\hbox{\scriptsize phys}}$,
as marked by the vertical dotted line in Fig.~\ref{Fig:MsqZ_almost_timelike}.
At this kink, both the propagator itself and the inverse propagator
functions have a branch-point, at which point the imaginary part
becomes nonzero.  With a nonzero exchange mass $\mu$, this kink occurs
well beyond the mass-pole at $p^2 = M^2(p^2)$, and both the propagator
and the inverse propagator functions are finite at this branch-point.
However, in the limit of $\mu \to 0$, this branch-point coincides with
the mass-pole singularity, as can be seen in the right panel of
Fig.~\ref{Fig:MsqZ_almost_timelike}.  Consequently, the propagator
exhibits a more complicated singularity instead of a simple mass-pole,
at which point the inverse propagator is zero, and a branch-cut starts
along the timelike axis.  The sign of the imaginary part is a
consequence of the $i\epsilon$ prescription -- or in the case of the
un-Wick rotation, of the direction of the rotation.

Due to the PV regularization, the (inverse) propagator has a second
kink along the timelike axis, located at
$(m_{\hbox{\scriptsize phys}} + \Lambda)^2$, beyond which the
imaginary parts fall off to zero, and the real parts of the (inverse)
fermion propagator approach their bare (tree-level) values,
see Fig.~\ref{Fig:AB_almost_timelike}.

\subsection{Spectral representation of the self-energy
  \label{sec:spectralrep}}
With PB regularization, the integral representation for the scalar and vector self-energies
can be written as
\begin{eqnarray}
  B(p^2) &=& m_0+\int_0^\infty ds \frac{\rho_B(s)}{p^2-s+i\varepsilon}\,
  \quad\text{with}\quad \rho_B(s)=-\text{Im}\left[B(s)/\pi\right]\, ,
  \label{rhob}\\
  A(p^2) &=& 1+\int_0^\infty ds \frac{\rho_A(s)}{p^2-s+i\varepsilon}\,
  \quad\quad\text{with}\quad \rho_A(s)=-\text{Im}\left[A(s)/\pi\right]\, ,\label{rhoa}
\end{eqnarray}
following the standard spectral representation of the
propagators~\cite{IZ}.  In principle, the spectral functions
$\rho_{A, B}$ fully determine the scalar and vector self-energies,
and thus the propagator.  

\begin{figure}[bth]
  \includegraphics[width=0.48\columnwidth]{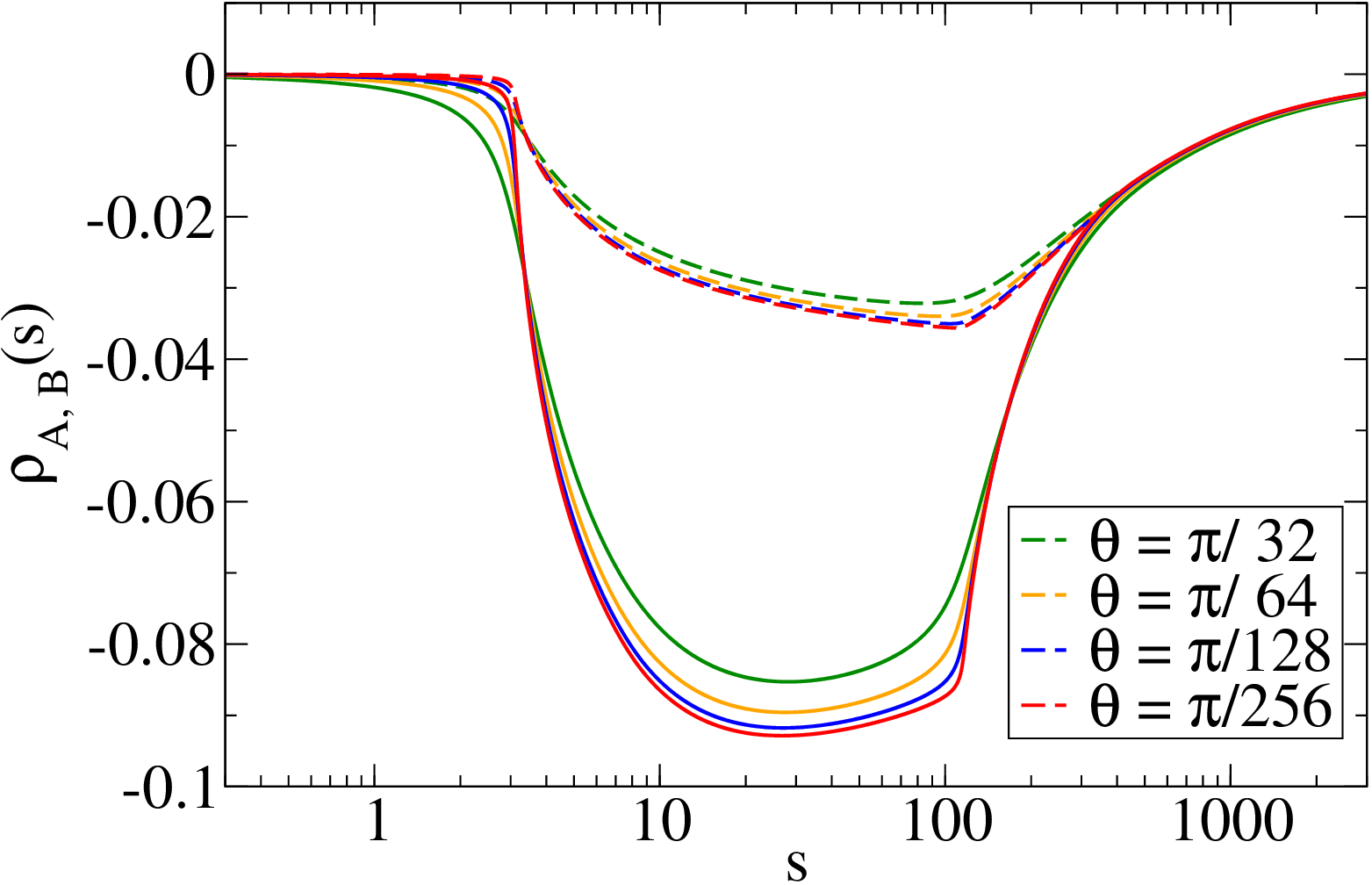}  \quad
  \includegraphics[width=0.48\columnwidth]{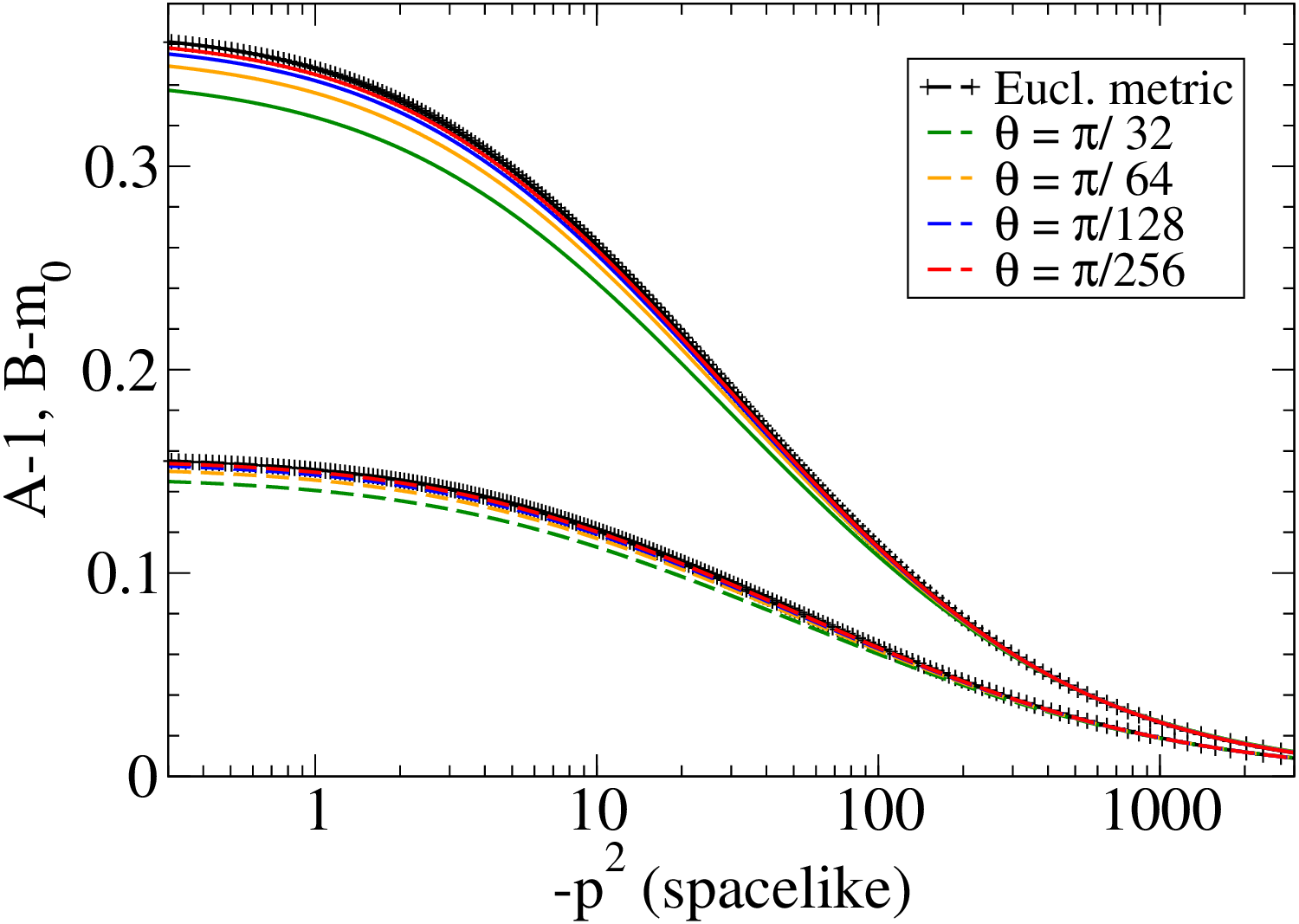} 
  \caption{ \label{Fig:check_spectral}
    Left: Approximate spectral functions $\rho_{A, B}$ obtained at
    different angles $\theta$ close to the timelike axis for
    $m_0 = 0.5$, $\mu=1.0$, $\Lambda = 10.0$, and $\alpha = 0.5$.
    Right: Spacelike self-energies obtained from the approximate
    spectral functions, compared to the Euclidean solution.  }
\end{figure}
In Fig.~\ref{Fig:check_spectral} we show on the left approximations to
the spectral functions $\rho_{A, B}$ obtained from the imaginary parts
of $A$ and $B$ at different angles $\theta$ close to the timelike
axis.  (Note that the angle $\theta$ is defined as the rotation angle
for $p_0$ or the magnitude of $p$; in terms of the variable $s$ used
in the spectral representation, this corresponds to an angle
$2\theta$.)  The right panel confirms that in the limit of $\theta \to
0$, these approximate spectral functions can indeed reproduce the
Euclidean (spacelike) to high accuracy.  With a more careful analysis
and using a Mellin transformation, we can use these `approximate
spectral representations' at nonzero values of $\theta$ to calculate
the self-energies in the entire slice of the upper complex $p^2$
plane, bounded by the real spacelike axis (negative $p^2$) and the
line $p^2 \, {\rm e}^{i\,2\theta}$.  More details will be presented in
Ref.~\cite{Maris:inprep}.

\section{Conclusion and outlook
  \label{sec:Conclusion}}
This contribution presents a preliminary study of the nonperturbative
fermion propagator in both the spacelike and near the timelike regions
by investigating the fermion DSE in rainbow-ladder truncation in the
Feynman gauge in a QED-like theory.  Two methods to solve the
Pauli--Villars regulated DSE were implemented to obtain the
self-energies near the timelike axis, both relying on an analytic
continuation of the Euclidean DSE into the complex momentum plane.  In
the first approach the energy component of the four-momenta are
complex-rotated to bring the Euclidean formulation towards the
Minkowski metric, while in the second method the magnitude of the
four-vector $p$ is complex-rotated to rotate the spacelike axis
towards the timelike axis.  Both methods were used to compute the
Dirac scalar and vector self-energies of the fermion near the timelike
region.  The second method showed to be much more accurate allowing
calculations with angles as small as $\theta=\pi/256 \approx
0.7^{\circ}$, quite close to the timelike axis.  This is natural as
with a fixed angle, in the first method the DSE has to be solved as
function of two real variables, while in the second approach the
scalar and vector self-energies depend on only one real variable,
allowing a finer grid in this one variable.

The coupling constant was chosen sufficiently large for the solutions
to allow for noticably nonperturbative effects, while below the value
for dynamical chiral symmetry breaking.  With a massive vector boson,
the obtained nonperturbative fermion propagator has a mass-pole at
$p^2 = M^2(p^2) = m^2_{\hbox{\scriptsize phys}}$ on the timelike
axis, followed by a branch-cut starting at
$p^2 = (m_{\hbox{\scriptsize phys}} + \mu)^2$.  With massless bosons,
$\mu=0$, this branch-cut starts at the physical mass, and the mass-pole
becomes a more complicated singularity.  Finally, the imaginary part
of the self-energies along the timelike axis were used to obtain the
spectral densities, from which the spacelike self-energies were
computed in good agreement with the Euclidean self-energies.

In the future, we intend to explore in more detail the analytic
structure of the fermion propagator in the complex plane by, e.g.,
generalizing the spectral representation with finite $\theta$
associated with the study the solutions of Laplace equations using
Mellin transform~\cite{Maris:inprep}; we also plan to extend these
investigations to other gauges, in particular the Landau gauge, and to
other theories.  The next step will be to use these nonperturbative
propagators in Minkowski metric for bound state calculations and to
explore hadron structure directly in Minkowski space.

\section*{Acknowledgments}
This work was supported by Funda\c c\~ao de Amparo \`a Pesquisa do
Estado de S\~ao Paulo, Brazil (FAPESP) Thematic grants No. 13/26258-4
and No. 17/05660-0, by CAPES, Brazil - Finance Code 001,
and by the US Department of Energy under Grants
No. DE-FG02-87ER40371 and No. DE-SC0018223 (SciDAC-4/NUCLEI).
TF thanks Conselho Nacional de Desenvolvimento Cient\'ifico e Tecnol\'ogico
(Brazil), Project INCT-FNA Proc. No. 464898/2014-5, and the Fullbright Visiting Professor Award.
DCD thanks FAPESP grant No. 17/26111-4.
EY thanks FAPESP grant No. 2016/25143-7.
PM thanks the Visiting Researcher Fellowship from FAPESP, grant No. 2017/19371-0.
This research used resources of the National Energy
Research Scientific Computing Center (NERSC), which is a US Department
of Energy Office of Science user facility, supported under Contracts
No. DE-AC02-05CH11231.



\begin{thebibliography}{99}

\bibitem{Eichmann:2016yit} 
  G.~Eichmann, H.~Sanchis-Alepuz, R.~Williams, R.~Alkofer and C.~S.~Fischer,
  Prog.\ Part.\ Nucl.\ Phys.\  {\bf 91}, 1 (2016)

\bibitem{Maris:1997tm} 
  P.~Maris and C.~D.~Roberts,
  Phys.\ Rev.\ C {\bf 56}, 3369 (1997)

\bibitem{Maris:1999nt} 
  P.~Maris and P.~C.~Tandy,
  Phys.\ Rev.\ C {\bf 60}, 055214 (1999)

\bibitem{Maris:2000sk} 
  P.~Maris and P.~C.~Tandy,
  Phys.\ Rev.\ C {\bf 62}, 055204 (2000)
  
\bibitem{Bhagwat:2006pu} 
  M.~S.~Bhagwat and P.~Maris,
  Phys.\ Rev.\ C {\bf 77}, 025203 (2008)
  
\bibitem{Maris:1997hd} 
  P.~Maris, C.~D.~Roberts and P.~C.~Tandy,
  Phys.\ Lett.\ B {\bf 420}, 267 (1998)

\bibitem{Atkinson:1978tk} 
  D.~Atkinson and D.~W.~E.~Blatt,
  Nucl.\ Phys.\ B {\bf 151}, 342 (1979).

\bibitem{Maris:1993PhD} 
  P.~Maris,
  {\it Nonperturbative analysis of the fermion propagator:
      Complex singularities and dynamical mass generation},
  Ph.D.~thesis, U. of Groningen (1993)

\bibitem{Maris:1994ux} 
  P.~Maris,
  Phys.\ Rev.\ D {\bf 50}, 4189 (1994).

\bibitem{NakaPTP69}  N. Nakanishi, Suppl. Prog. Theor. Phys. {\bf 43}, 1 (1969)

\bibitem{Sauli:2002tk} 
  V.~Sauli,
  JHEP {\bf 0302}, 001 (2003)

\bibitem{Sauli:2004bx} 
  V.~Sauli,
  Few Body Syst.\  {\bf 39}, 45 (2006)

\bibitem{Sauli:2006ba} 
  V.~Sauli, J.~Adam, Jr. and P.~Bicudo,
  Phys.\ Rev.\ D {\bf 75}, 087701 (2007)
  
\bibitem{Jia:2017niz} 
  S.~Jia and M.~R.~Pennington,
  Phys.\ Rev.\ D {\bf 96},  036021 (2017)

\bibitem{Kusaka:1995za} 
  K.~Kusaka and A.~G.~Williams,
  Phys.\ Rev.\ D {\bf 51}, 7026 (1995)

\bibitem{Kusaka:1997xd} 
  K.~Kusaka, K.~M.~Simpson and A.~G.~Williams,
  Phys.\ Rev.\ D {\bf 56}, 5071 (1997)

\bibitem{Karmanov:2005nv} 
  V.~A.~Karmanov and J.~Carbonell,
  Eur.\ Phys.\ J.\ A {\bf 27}, 1 (2006)
   
\bibitem{Sauli:2010ava} 
  V.~Sauli,
  Few Body Syst.\  {\bf 49}, 223 (2010)

\bibitem{Frederico:2011ws} 
  T.~Frederico, G.~Salm\`e and M.~Viviani,
  Phys.\ Rev.\ D {\bf 85}, 036009 (2012)

\bibitem{Frederico:2013vga} 
  T.~Frederico, G.~Salm\`e and M.~Viviani,
  Phys.\ Rev.\ D {\bf 89}, 016010 (2014)

\bibitem{Carbonell:2010zw} 
  J.~Carbonell and V.~A.~Karmanov,
  Eur.\ Phys.\ J.\ A {\bf 46}, 387 (2010)
  
\bibitem{dePaula:2016oct} 
  W.~de Paula, T.~Frederico, G.~Salm\`e and M.~Viviani,
  Phys.\ Rev.\ D {\bf 94}, 071901 (2016)

\bibitem{dePaula:2017ikc} 
  W.~de Paula, T.~Frederico, G.~Salm\`e, M.~Viviani and R.~Pimentel,
  Eur.\ Phys.\ J.\ C {\bf 77}, 764 (2017)

\bibitem{Castro:2019tlh} 
  A.~Castro, E.~Ydrefors, W.~de Paula, T.~Frederico, J.~H.~De Alvarenga Nogueira and P.~Maris,
  arXiv:1901.04266 [hep-ph], to appear in the proceedings of XLI RTFNB, Maresias, Brazil, Sept. 2018

\bibitem{IZ} C.~Itzykson and J.-B.~Zuber, {\em Quantum Field Theory}, McGraw-Hill, New York, 1985.

\bibitem{Maris:inprep}
  P.~Maris, S.~Jia, D.~C.~Duarte, W.~de Paula, E.~Ydrefors, and T.~Frederico, in preparation (2019).
  
\end{thebibliography}
\end{document}